\documentclass[11pt]{article} 

\usepackage[utf8]{inputenc} 
\usepackage{geometry} 
\usepackage{graphicx} 
\usepackage{booktabs} 
\usepackage{array} 
\usepackage{amsmath}
\usepackage{verbatim} 
\usepackage{hyperref}

\begin{document}

\title{CITY-SCALE SIMULATION OF COVID-19 PANDEMIC \& INTERVENTION POLICIES USING AGENT-BASED MODELLING}

% AUTHOR: Enter the authors of the article, see end of the example document for further examples
\author{Gaurav Suryawanshi\\[12pt]
Department of Mathematics\\
Indian Institute of Technology Kharagpur\\
Kharagpur, INDIA, 721302\\
% Multiple authors are entered as follows.
% You may also need to adjust the titlevbox size in the preamble - search for titlevboxsize
\and
Varun Madhavan\\[12pt]
Department of Chemical Engineering\\
Indian Institute of Technology Kharagpur\\
Kharagpur, INDIA, 721302\\
\and
Adway Mitra\\ [12pt]
Centre of Excellence in Artificial Intelligence\\
Indian Institute of Technology Kharagpur\\
Kharagpur, INDIA, 721302\\
\and
Partha Pratim Chakrabarti\\ [12pt]
Centre of Excellence in Artificial Intelligence, \\
Department of Computer Science and Engineering\\
Indian Institute of Technology Kharagpur\\
Kharagpur, INDIA, 721302\\
}

\maketitle

\section*{ABSTRACT}
During the Covid-19 pandemic, most governments across the world imposed policies like lock-down of public spaces and restrictions on people's movements to minimize the spread of the virus through physical contact. However, such policies have grave social and economic costs, and so it is important to pre-assess their impacts. In this work we aim to visualize the dynamics of the pandemic in a city under different intervention policies, by simulating the behavior of the residents. We develop a very detailed agent-based model for a city, including its residents, physical and social spaces like homes, marketplaces, workplaces, schools/colleges etc. We parameterize our model for Kolkata city in India using ward-level demographic and civic data. We demonstrate that under appropriate choice of parameters, our model is able to reproduce the observed dynamics of the Covid-19 pandemic in Kolkata, and also indicate the counter-factual outcomes of alternative intervention policies.
%This set of instructions for producing a proceedings paper for the 2020 Winter Simulation Conference (WSC) with \LaTeX\ also serves as a sample file that you can edit to produce your submission, and a checklist to ensure that your submission meets the WSC 2020 requirements. Please follow the guidelines herein when preparing your paper. Failure to do so may result in a paper being rejected, returned for appropriate revision, or edited without your knowledge.

\section{INTRODUCTION}\label{sec:intro}
The Covid-19 pandemic, caused by the SARS-COV-2 virus, broke out from Wuhan in China in December 2019, and has spread to almost every country of the world. Till March 2021, nearly 120 million people have been affected so far, with over 2 million deaths. Among the countries most severely impacted in terms of total caseload and total number of deaths is India, which has a vast population of nearly 1.4 billion and moderately high population density of 382 persons per Sq Km. The large and populous cities of India have faced high caseload. One of the largest and most densely populated cities in India is Kolkata, located in the eastern part, which has had its share of cases in the pandemic. 

In the absence of medicines or vaccines specifically equipped to tackle this disease, most national, regional and local governments across the world have resorted to non-pharmaceutical interventions (NPI), such as prescription of face-masks and hand sanitizers, and maintenance of ``physical distancing" through stay-at-home orders, partial or total closure of public spaces like schools and offices, and so on. While such measures may be useful in slowing down the pandemic spread, they cannot be sustained long due to their disruptive effects on society and economy. This disruption is particularly severe in places like India where a large part of the population earn their livelihood through daily wages. Governments have to make a trade-off between public health and economics while implementing policies on NPIs. This necessitates the assessment of policy impacts, which may be achieved through computer simulations of the societal impacts of the pandemic as well as various intervention policies.

Traditionally, pandemics have been simulated using epidemiological models. However, such models usually operate at very course scales by considering gross statistics such as number of infected people, and cannot take into account spatial aspects of the pandemic spreading process. To simulate the pandemic spreading process under real or hypothetical policy interventions, it is necessary to understand the daily lifestyles and social contacts of individuals and the associated spatial structures. It is here that agent-based models become important, as they can provide a representation of individual behavior at high spatial and temporal resolution. The models can take into account both spatial and social structures in a limited region, such as a city. An agent-based model can be used to emulate the daily activities of individuals under the influence of the pandemic and the NPIs, and observe the resultant impacts. Once the model is validated against actual observations, it can be used to simulate counterfactual or ``what-if" scenarios, such as alternative NPI policies, and assess their impacts. During the Covid-19 pandemic, several research groups all over the world have been developing agent-based models for specific cities, or larger regions~\cite{cov_6}. Different models have different levels of detailing with respect to the region of study, and may include any subset of features such as demographic factors, family structures, spatial layout, residential areas and public spaces, transport systems, daily schedule of people and so on. 

This paper presents another agent-based model for simulating the pandemic in cities. To the best of our knowledge, our model has the highest level of detailing, and we explicitly emulate all the features mentioned above. We also have additional features such as compliance rates, i.e. the probability that any individual will obey the operational NPI. In this study, we have collected the parameters specifically for the city of Kolkata, and validated the model by reproducing the time-series of daily infections of Covid-19 as observed in the records. We have demonstrated that explicitly emulating the aforementioned city structures in the model improves this validation. Additionally, we have also analyzed a number of ``what-if" scenarios about alternative NPI policies such as no lockdown, localized lockdowns in hotspots, and partial closing of workplaces.

\section{RELATED WORK}
In this section, we review the relevant literature on agent-based models and their applications to modeling cities and pandemic spreading.

\textbf{Agent-based Simulation:} Agent-based modeling (ABM), as a computational technique, emerged in the 1990s, primarily in the domain of computational social science~\cite{abm_1,abm_2} as a testbed for artificial societies. It was also used in the domains of economics~\cite{abm_5}, ecology~\cite{abm_4} and population biology~\cite{abm_6}. Its potential for modeling urban systems was noted by~\cite{abm_3} to capture various stakeholders, urban components and their interactions, such as the work by~\cite{ca_urban},~\cite{ca_abm} that explored the temporal changes in urban sprawl and land-use using cellular automata-based and agent-based models comparatively. Agents were used to represent individual residents in cities and rule-based modeling of their interactions was developed~\cite{ct_1,ct_2}. More recent urban ABMs focus on specific aspects of cities in greater details, such as residential areas ~\cite{rs_1}, and transport~\cite{tr_1,tr_2,tr_3}.  MATSIM~\cite{MATSIM} and UrbanSim~\cite{ct_3} are ABM-based software packages to simulate long-term trends in a city related to traffic management, electricity/water usage and gas emissions etc, for use in multi-dimensional policy planning. ~\cite{plc_1} focus on the intersection between urban mobility, equitable accessibility and air pollution using ABMs. 

\textbf{Epidemic Model:} Epidemiology, or the dynamics of epidemics have traditionally been studied with differential equation-based compartmental models such as Susceptible-Infected-Recovered (SIR). However, such models deal with only gross statistics of pandemic such as number of susceptible or infected people, without a realistic representation of how infections may spread from person to person depending on their locations and interactions. So modern epidemiology is turning towards ABMs, such as~\cite{epi_1,epi_2,epi_3,cov_4}. Such models need i) a synthetic population with realistic demographic attributes, ii) a geo-located social network of the population, and iii) a disease progression and infection model. The spread of any epidemic in an urban area is strongly dependent on the transportation systems, and this is captured in the recent study by~\cite{epi_4} using agent-based models and MATSIM software for transport systems. Other examples of agent-based models using networks are~\cite{cov_7,cov_10}.

During the Covid-19 pandemic, agent-based models have acquired special attention worldwide, as policymakers are interested to assess the impacts of various non-pharmaceutical interventions (NPI). Such impacts may be desirable (slowing down of pandemic spread) or undesirable (economic disruption). The study~\cite{cov_1} proposes COVID-ABS model to study the impact of various NPIs on both public health and economics by emulating a closed society including its people and decision-makers. This framework uses SEIR-type disease model and considers demographic and economic attributes and daily schedules of people. The parameters of this model have been chosen based on Brazil, but can also be used for any other place. A similar study~\cite{cov_3} was done for France, using detailed attributes related to demographics and social contact structure of the agent population, and a small statistical model for city spaces and transport. Another similar model was built for cities and validated on Ford County, Kansas, USA using mostly demographic attributes of the agents, along with smartphone-based contact tracing. Another study~\cite{cov_2} was built specifically for small cities, taking into account the basic structure of the city, population demographics, testing and contact tracing, and calibrated with New Rochelle, New York. This model also includes confounding due to covid-like symptoms and vaccination, along with a detailed scenario analysis. Several other studies like~\cite{cov_9,cov_8} also present similar models validated on particular cities or regions.

Based on the above survey, we identify several features related to modeling Covid-19 pandemic in cities. Each of the studies takes into account a subset of these features. In Table~\ref{tab:model_comparison}, we list out these features, showing which study takes into account each feature elaborately with high (H), medium (M) or low (L) model complexity. High model complexity means that there are separate variables with attached stochastic processes representing the feature, medium complexity means that it is managed with proxy parameterization, and low complexity means that the feature is not represented at all. These features include the disease model, demographic attributes and their daily schedule related to agents, the spatial layout of the city with locations of residences, workplaces etc, the transport system and medical facilities of the city. Other features include economic impacts, vaccination, and the presence of confounding diseases with similar symptoms. Other important features of these models are their analysis of various hypothetical scenarios related to intervention policies, and validation of the models by demonstrating their ability to reproduce the actual observations related to number of daily cases, deaths etc in the target location.

\begin{table}[]
    \centering
    \caption{Comparison of different agent-based models for Covid-related scenario analysis, in terms of urban features (H-High, M-Medium, L-Low) considered by them}
    \begin{tabular}{|c|c|c|c|c|c|c|}
    \hline
    Features & Silva et al & Hoertel et al & Truszkowska & Shamil et al & Kerr et al & This Work\\
    & & & et al & & &\\
    \hline
Disease Model           & H  &  H  &  H  &  H & H & H\\
Demographic attributes  & H  &  H  &  H  &  H & H & H\\
Daily Schedule          & H  &  M  &  L  &  L & L & H\\
City Structure          & L  &  M  &  H  &  L & L & H\\
Workplaces              & M  &  M  &  H  &  L & M & H\\
Transport Systems       & L  &  M  &  L  &  L & L & M\\
Medical Facilities      & L  &  L  &  M  &  L & L & M\\
Contact Tracing         & L  &  L  &  H  &  M & M & H\\
Economic Impacts        & H  &  L  &  L  &  L & L & L\\
Confounding             & L  &  L  &  H  &  L & L &L\\
Vaccination             & L  &  L  &  H  &  L & L & L\\
Scenario analysis       & H  &  H  &  H  &  H & H & H\\
Validation              & M  &  H  &  H  &  H & H & H\\
    \hline
    \end{tabular}
    
    \label{tab:model_comparison}
\end{table}

\section{MODEL COMPONENTS}
\subsection{Agents}
The agents of this model are the individuals, i.e. the residents of the city. Each agent is associated with attributes shown in Table \ref{tab:agents}. 

\subsection{City Model}
A city contains many kinds of spaces - residential, workplaces, marketplaces etc, as well as different kinds of facilities and services. In this work, we consider statistical models for the following.

\textbf{Centres of Economic activity or Workplaces:}
The economic activities of the city are driven by several industries. The economic sphere is divided into sectors, which are further divided into sub-sectors. Each sub-sector then contains a number of workplaces under it. Workplaces are locations where agents go for work. Each workplace is assigned with location co-ordinates. Also, agents representing working individuals are associated with a workplace. Each workplace holds two kinds of people, one is one-time visitor and others are daily wage workers. A workplace has attributes as shown in Table \ref{tab:workplace}. Further, the workplaces in different sectors (education, healthcare, hotels etc)  have different characteristics related to the exposure and physical interactions among the workers, which has a bearing on the infection spread. These characteristics are detailed in Table \ref{tab:workers}, with reference to Kolkata city.

\textbf{Transportation:} There are two kind of transportation, inter-city and intra-city. Intra-city transport enables citizens to travel from one place to another within the city, especially home to workplace. Although such transport can be either public or private, we ignore private transportation in this model since disease infection in a private vehicle is highly unlikely. 

\textbf{Healthcare:} The crucial healthcare sector consists of a network of COVID-19 Hospitals, COVID-19 Healthcare centres, COVID Isolation Centres. Each of these facilities are associated with a location, and has a fixed capacity. The attributes of healthcare facilities are mentioned in Table \ref{tab:workplace}.

\textbf{Education:} In this model, education has been treated separately from the economic sectors, due to the extremely high number of students. These students may be interpreted as an employed workforce generating negligible economic output. Colleges and places of higher education are also under the Education sector. Each educational institute too is associated with a location and capacity. Each agent representing an individual who is a student is assigned to an educational institute (school or college), depending on age. 
\begin{table}[ht]
    \begin{minipage}[b]{80mm}
    %\centering
    \caption{Agent attributes.\label{tab:agents}}
    \begin{tabular}{p{0.20\linewidth} | p{0.1\linewidth} | p{0.50\linewidth}}
    \textbf{Property} & \textbf{Basis} & \textbf{Description} \\ 
    \hline
    ID & $Z$ & ID of the person \\
    Age & $Z$ & Age of the person \\
    Age Group & $Z$ & Bands of 10 years\\
    Family ID& $Z$ & Assigned to a family\\
    isCitizen & \{0, 1\} & resident of the city or not\\
    Workplace & $Z^{3}$ & 3-D vector (Sector, Sub-Sector, Workplace ID) \\ 
    Visiting Places & $Z^{n}$ & $n$-D vector - frequently visited places\\ 
    Comorbidity & \{0, 1\} & has co-morbidity or not \\
    Income level & $R$ & Family daily income\\ 
    Occupancy & $Z$ & \\
    \end{tabular}
    \end{minipage}
    \begin{minipage}[b]{80mm}
    \caption{Workplace attributes.\label{tab:workplace}}
    \begin{tabular}{p{0.20\linewidth} | p{0.1\linewidth} | p{0.50\linewidth}}
    \textbf{Property} & \textbf{Basis} & \textbf{Description} \\ 
    \hline
    ID & $Z$ & Id of the workplace \\
    Sector & $Z$ & Economic Sector to which workplace belongs\\
    Sub-Sector & $Z$ & Sub-sectors within sector\\
    Location & $R^{2}$ & (Latitude, longitude) of the workplace \\
    isEssential & \{0, 1\} & whether the workplace is essential\\
    Workers & $Z^{n}$ & $n$-D vector, list of people working\\ 
    Visitors & $Z^{n}$ & $n$-D vector, list of people visiting \\
    Income level & $R$ & Daily economic output of the Workplace \\
    \end{tabular}
    %\centering
    \end{minipage}
\end{table}

\subsection{Infectious Model}
Now we come to the most crucial part- the Covid-19 infection model. For this model, we define the following \emph{states}, such that each agent is associated with one state at every time-step.

Each agent consists of a state belonging to a finite state machine, which is a vector of dimension 2. First dimension represents the virus-related state  i.e Healthy (H), Infected (IF) and Recovered (R) and Dead (D). Infected state has two sub-parts namely Symptomatic (S) and Asymptomatic (A). We call them Virus-State (\(S_v\)). Second dimension represents mobility-related state which are Free (F), Out-of-City (O), Quarantined (Q), Isolated (I) and Hospitalized (HP). We call them Mobility-State (\(S_m\)) These two kinds of states are independent of each other, and the final state consists of a concatenated vector of these two.

Regarding the Virus-state, initially every person is assumed to be healthy (H). When a person in state $H$ comes in contact with a person in state $IF$, the former gets infected with a certain probability. The viral load (VL) of a newly infected person is assumed to follow a Beta distribution, scaled appropriately. If this load is below a threshold, the person remain asymptomatic. Otherwise, the person undergoes a fixed incubation period after infection, after which they start showing symptoms. The peak infection period is sampled from a Gaussian distribution, after which the person may die or start recovering (depending on age and co-morbidity). The time taken to recover is also considered to follow a Gaussian distribution.

Infection can spread from one person to another with a certain probability whenever they share the same physical space. This probability is a function of the contact duration and physical distance between them. This is why we have considered number of work-hours and physical gap as attributes of workplaces.

\subsection{Simulation Methodology}
Having described the model in details, we now come to the simulation part. The simulation proceeds by initializing the system, which includes setting up all the agents and their attributes, workplaces, medical facilities etc. This is done by sampling from probability distributions, such that all the attributes are consistent with each other. We then set the hyper-parameters of the model such as parameters of various distributions in section 3.3. Additionally we define Compliance Rate: the proportion of population actually following the policy in place. Higher compliance rates means no two agents meet beyond their daily schedule.  A small random subset of the population is infected with the disease as initialization. After that, we begin the simulation for the specified period. At each time-step we sample the movements of each agent according to a stochastic process based on their daily routine, whose distributions are based on the agent's workplace, residence and travel preferences. We track the interactions between pairs of individuals (when they come spatially close during their movements), and in such cases infection may take place stochastically as explained above. 
% \begin{comment}

% \end{comment}

% \begin{comment}
\begin{table}[ht]
    \begin{minipage}[b]{70mm}
    
    \centering
    \caption{Healthcare workplace attributes.\label{tab:health}}
    \begin{tabular}{p{0.20\linewidth} | p{0.1\linewidth} | p{0.50\linewidth}}
    \hline
    \textbf{Property} & \textbf{Basis} & \textbf{Description} \\ 
    \hline
    ID & $\mathcal{Z}$ & Id of the workplace \\
    Number of beds & $\mathcal{Z}$ & Number of beds that the
    place is able to host \\
    Workers & $\mathcal{Z^n}$ & Workers working at the hospital \\
    ICU beds & $\mathcal{Z}$ & Numbers if beds at the intensive care facility \\
    Ventilators & $\mathcal{Z}$ & Crucial resource whose availability can affect death rates \\
    Type & \{Free, Paid\} & Whether the hospital is free for everyone or one has to pay \\ 
    \hline
    \end{tabular}
    
    \end{minipage}
    \begin{minipage}[b]{120mm}
        % \centering
        %\hspace{-0.5mm}
        \caption{Workers per sector.\label{tab:workers}}
        \begin{tabular}{p{0.2\linewidth} | p{0.1\linewidth} | p{0.1\linewidth} | p{0.1\linewidth} | p{0.1\linewidth}}
        \hline
          \textbf{Sector} & \textbf{No. of worker} & \textbf{No. of centers} & \textbf{working hours} & \textbf{Physical gap} \\ \hline
            Education &       720801 &  13900 &         8 &     0.5\\
            Commerce &        45329 &   120 &           8 &     3\\
            Healthcare &      22634 &   [16, 22, 19] &  0 &   2\\
            Agriculture &     1512 &    170 &           8 &     5\\  
            Manufacturing &   136549 &  450 &           8 &     2\\ 
            Mining &          20 &      3 &             8 &     1\\
            Utilities &       22202 &   55 &            8 &     2\\    
            Construction &    444782 &  1472 &          8 &     5\\
            Hotel &           962881 &  4162 &          8 &     2\\
            Finance &         626412 &  2068 &          8 &     3\\
            Social &          341858 &  1792 &          12 &    1\\
        \hline
        \end{tabular}
        
    \end{minipage}

\end{table}

% \end{comment}
\section{RESULTS FOR KOLKATA CITY}
We parameterized the above simulation framework according to Kolkata, a city in Eastern India, as a test case. We initialized it by injecting actual data related to this city as the model's parameters and attributes. 

\subsection{City-scale Data for Kolkata}
To set the attributes and parameters for the simulation in Kolkata, we utilized the following sources of information:

\textbf{Demographics}: Distribution of population across the wards of the city, including the number of people in each ward and the population density of each ward are available at: \url{https://censusindia.gov.in/pca/pcadata/pca.html},\url{https://censusindia.gov.in/2011census/population_enumeration.html}

\textbf{Medical facilities}: Distribution of Covid Quarantine Centres, Covid Care Centres and Covid Hospitals across the wards are available at
    \url{https://www.wbhealth.gov.in/uploaded_files/corona/Notification___Revised___67_Covid_Hospital___30.04_.2020_.pdf}, \url{https://www.wbhealth.gov.in/uploaded_files/corona/Covid-19_Pvt_Hospitals_Kol_Isolation_22_March_20.pdf}
 
\textbf{Covid testing facilities}: The Covid-19 testing capacity in Kolkata are available at \url{https://www.wbhealth.gov.in/uploaded_files/corona/Order_157_05052020_For_attention_of_the_Public_SB_Secy.pdf} 
    
\textbf{Educational centres} Distribution of schools and colleges across the wards are available at \url{http://www.wbpublibnet.gov.in/sites/default/files/sites/default/files/sites/default/files/uploads/pdf/school_list_kolkata.pdf}, \url{https://wbhed.gov.in/page/govt_colleges.php}

\begin{figure}
  \begin{minipage}{.5\textwidth}
    \centering
    \includegraphics[width = \textwidth, height = 2.2in]{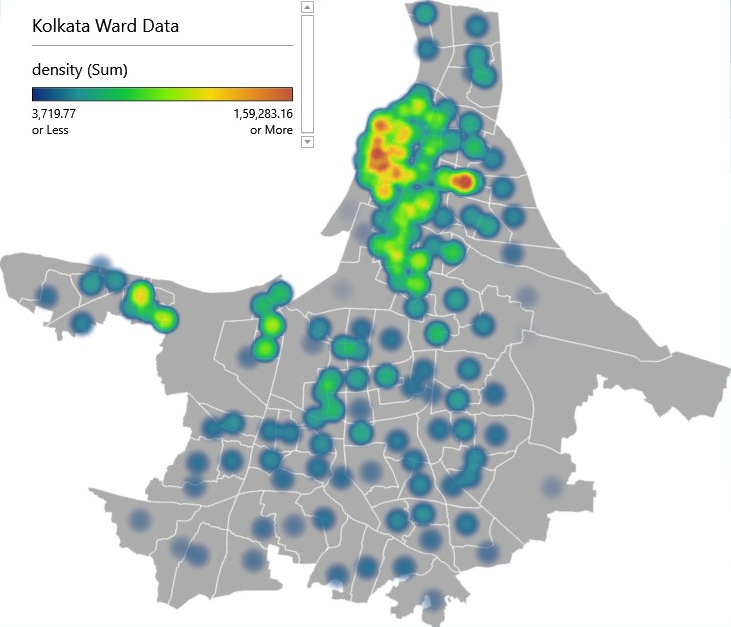}
    \caption{Population density in Kolkata city}
  \end{minipage}
  \begin{minipage}{.5\textwidth}
    \centering
    \includegraphics[width = \textwidth, height = 2.2in]{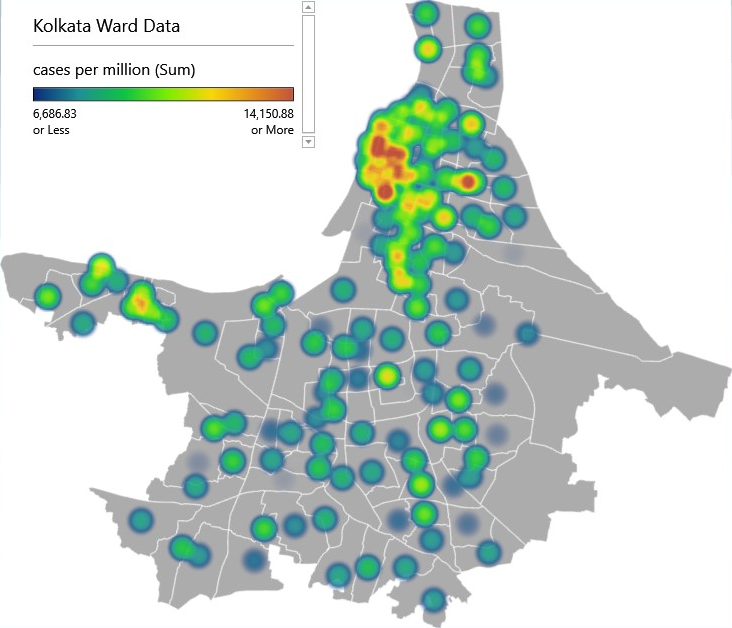}
    \caption{Baseline positive cases per million}
  \end{minipage}

\end{figure}

\subsection{Model Simulations}
The simulation was started from May 3rd, 2020. This is the date from which we have the Kolkata district level data. Although entire India went into lockdown since 25th March 2020, all COVID-19 related data was published state-wise till 3rd May. Since there is a gap between data availability and start of pandemic, we need to specify the number of infected people till this date. For this purpose, we used reverse simulation, i.e. start the simulation from March 25 with various random initializations and identify the one for which the "Positive Tested" count on May 3 matches the recorded data (651) for the same day. The number of active cases (1495) on that day was then obtained from that simulation.

The lockdown policy, containment zone policy, testing capacity and public transport capacity were set precisely as actually followed in the city of Kolkata. We used publicly available data \href{ https://api.covid19india.org/}{Covid-19 India API} for the same. The education sector (schools and colleges) in Kolkata can be a major source of infections due to high density of students/teachers. Hence the state government kept all schools and colleges shut through the year. The same is assumed in the simulations.

To fit the predicted cases to the actual number of cases, we used grid search on parameters such as Compliance Rate and IFP values during each month, upto the month of October. The compliance rate with the policies mentioned above were chosen to be 0.5 during May 1 - May 14, 0.8 during May 15 - July 31, and 0.4 from August 1 onwards. This was done to mimic tendency of poeple to follow the rules. The effect of varying these values are discussed in Section 5. The probability of infections (IFP) via external transportation routes was chosen to be 0.01 in June (as most transport services were closed), which was increased gradually to 0.25 in September. Contact Tracing efficacy was chosen to be 60\% for workplace contacts and 30\% for transport contacts. 

The simulation was carried out with the above model, along with the parameter settings mentioned above for the period May 3 to 1 October. For each parameter setting, the simulation was run 10 times, and the mean values reported. The number of new daily infections as obtained by the model are plotted in Fig 4(ward distributed) against the number of new daily infections as reported \href{ https://api.covid19india.org/}{here}. In 108 out of 152 days, simulation error remains within the 10\% . Most of the errors related to prediction occur during the early parts of the simulation which is expected, as the pandemic had started well before the simulation start date as mentioned in section 4.2. In Figures 1-3 we can see the 141 wards of the city, and the simulated number positive cases per ward. We see a strong correlation between the densities of Ward and cases, indicating a confirmation on trends we have seen globally.

\subsection{Simulations with Simplified Model}
As a possible alternative to using this fine-grained ward-level data, we attempted to recreate the simulation with a simplified version of the model which assumes uniform distribution of all parameters, i.e. population, population density, density of hospitals etc, across all wards leaving rest of the hyper-parameters as same. We observe in Figure 4 that this simulation does not perform as well as our proposed model's simulation, as the daily error is higher on most days. This suggests that high spatial granularity is important for successful simulation of pandemic in cities.

\begin{figure}[h]
    \centering
    \includegraphics[width = 1\textwidth, height = 5cm]{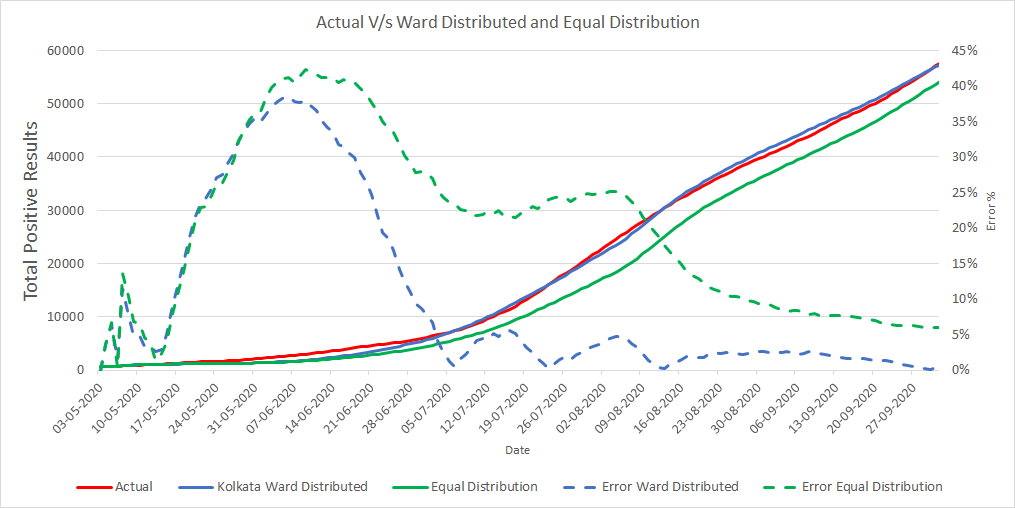}
    \caption{Comparison with reduced model, here Actual represents the Ground truth, Kolkata Ward Distributed represents the simulation with true ward population distributions and Equal Distribution represents the simulation with equally distributed ward population.}
\end{figure}

\section{"WHAT-IF" ANALYSIS OF COUNTERFACTUAL SCENARIOS}
The real aim of this kind of simulation is to generate and analyze counterfactual scenarios, such as what could have been possible results of alternative policies had they been implemented. Using our agent-based simulator, we predict the outcome of various scenarios (i.e. intervention policies) by adjusting the parameters accordingly.
\begin{itemize}
    \item \textbf{Lockdown Phases:} We consider the following lockdown parameters - %the extent and duration of lockdowns, including fixed duration lockdowns at the start of the pandemic and Containment Zone lockdowns, where specific wards with a spike in the number of cases were locked down. 
        \begin{itemize}
            \item \textbf{Start date of lockdown} - date when the lockdown comes into effect
            \item \textbf{Duration of lockdown} - the number of days for which the lockdown is in effect
            \item \textbf{Type of lockdown} - the lockdown can be either city-wide, or limited to specific wards or zones with high number of cases.
        \end{itemize}
    
    \item \textbf{Use of Public Transport:} This is the fraction  of the total population who are expected to use public transport on daily basis. In Kolkata city, this ratio is estimated to be about 0.17.
%        \begin{equation*}
%            \begin{split}
%                useTN = \frac{E(A_{TD}=1)}{P}
%            \end{split}
%        \end{equation*}
%        \textit{$A_{TD}$} is equal to 1 if the agent uses public transport on a daily basis
    \item \textbf{Extent of Physical Distancing:} Transmission probability between two persons depends on the distance between them. 
    \item \textbf{Contact tracing efficacy:} whenever any person tests as positive, what fraction of their close contacts are identified for testing?
    \item \textbf{Compliance Rates:} the effect of choosing a range of Compliance Rates
\end{itemize}

\subsection{Scenarios Evaluated}

%\begin{figure}
%    \begin{minipage}{.5\textwidth}
%        \centering
%        \includegraphics[width = \textwidth, height = 1.8 in]{scenarios/no_lockdown.png}
%        \caption{No lockdown.\label{fig: no lock}}
%    \end{minipage}
%    \begin{minipage}{.5\textwidth}
%        \centering
%        \includegraphics[width = \textwidth, height = 1.8 in]{scenarios/only_education_locked.png}
%        \caption{Only Education sector locked\label{fig: only edu lock}}
%    \end{minipage}
%\end{figure}

\begin{figure}
  \begin{minipage}{.5\textwidth}
    \centering
    \includegraphics[width = \textwidth, height = 2 in]{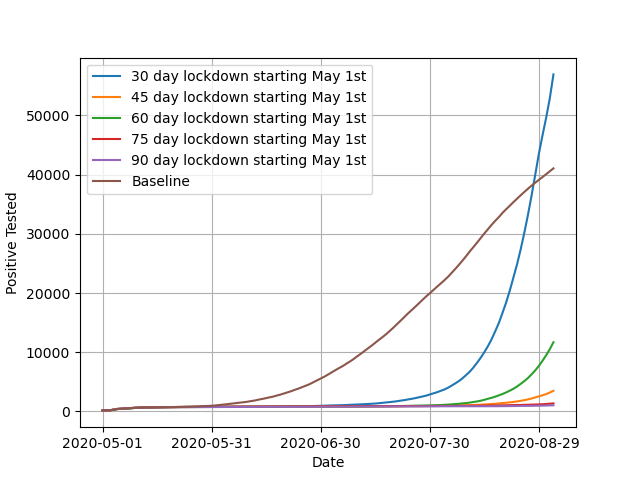}
    \caption{Long lockdown.\label{fig: long lock}}
  \end{minipage}
  \begin{minipage}{.5\textwidth}
    \centering
    \includegraphics[width = \textwidth, height = 2 in]{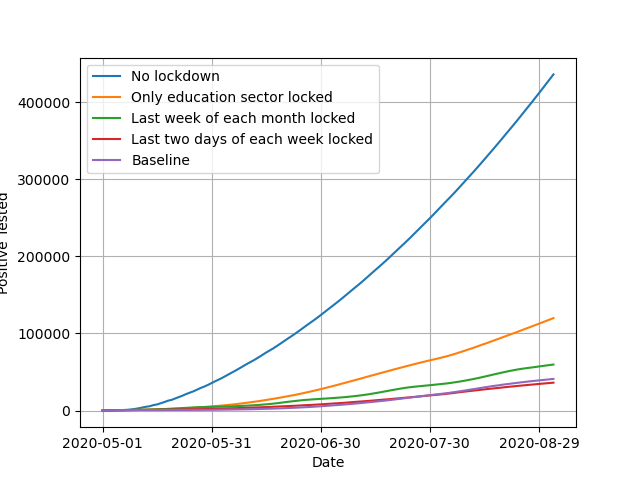}
    \caption{No Lockdown + Only Education Sector Locked + Multiple Short Lockdowns.\label{fig: multi}}
  \end{minipage}
\end{figure}
\begin{figure}
  \begin{minipage}{.5\textwidth}
    \centering
    \includegraphics[width = \textwidth, height = 2 in]{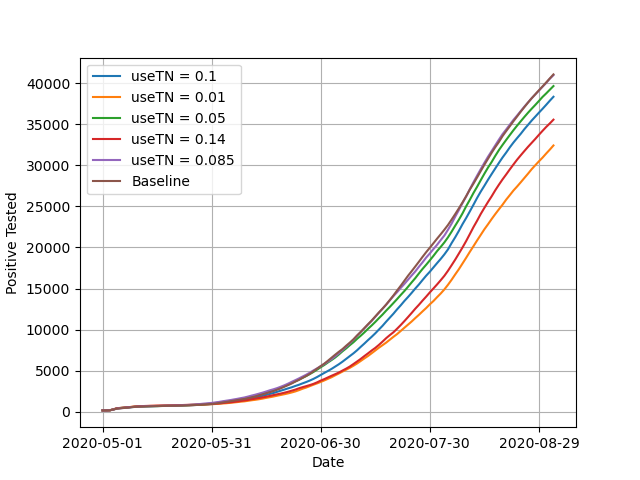}
    \caption{Effect of reduced public transport usage.\label{fig: transport}}
  \end{minipage}
  \begin{minipage}{.5\textwidth}
    \centering
    \includegraphics[width = \textwidth, height = 2 in]{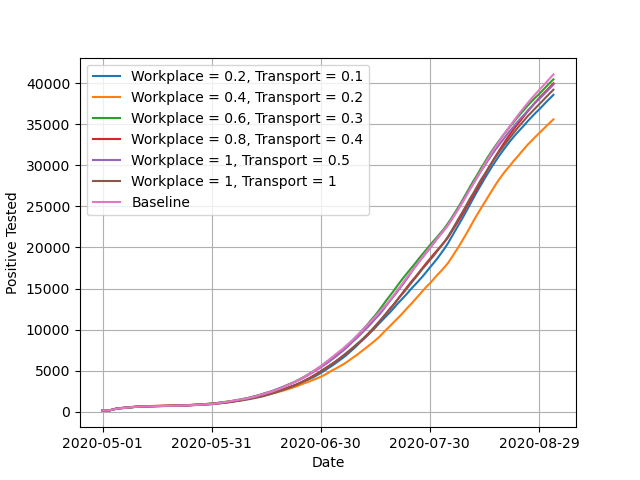}
    \caption{Contact Tracing Efficacy.\label{fig: contact tracing}}
  \end{minipage}
\end{figure}
\begin{figure}
  \begin{minipage}{.5\textwidth}
    \centering
    \includegraphics[width = \textwidth, height = 2 in]{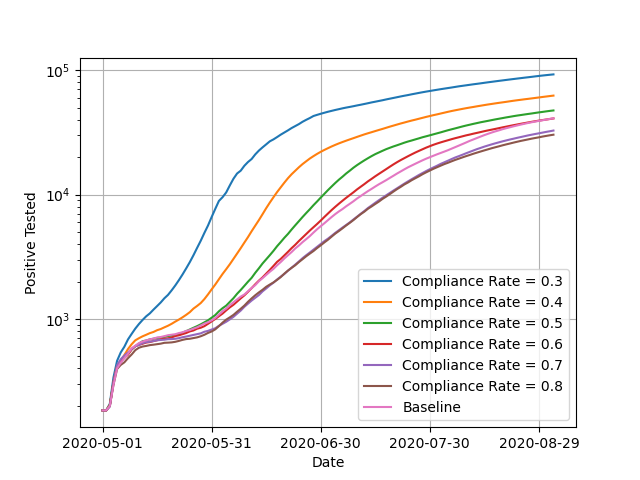}
    \caption{Effect of compliance rate.\label{fig: comp rate}}
  \end{minipage}
  \begin{minipage}{.5\textwidth}
    \centering
    \includegraphics[width = \textwidth, height = 2 in]{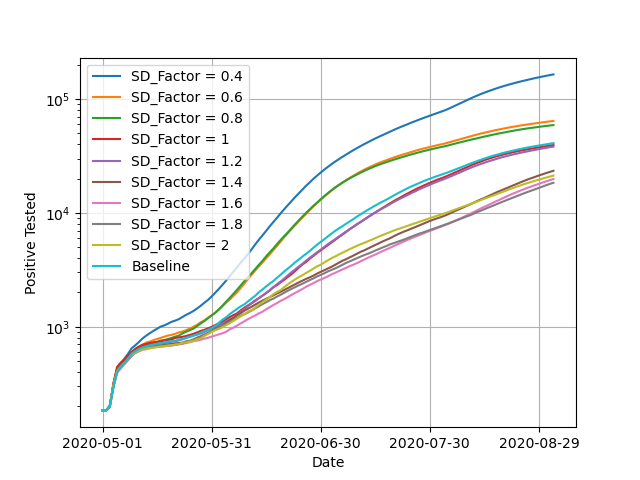}
    \caption{Effect of social distancing.\label{fig: social dist}}
  \end{minipage}
\end{figure}

\begin{itemize}
    \item \textbf{Lockdown Variations}
        \begin{itemize}

            \item \textbf{Long lockdown:} Lockdowns of 30, 45, 60, 75 and 90 days were simulated from May 1st. This scenario was tested to see whether the virus could be completely exterminated with a long lockdown. Out-of- city travel was completely stopped and the Education sector was kept locked throughout the simulation. The results are shown in Figure \ref{fig: long lock}. We find that prolonging the lockdown does not help as cases shoot up later.
            
            \item \textbf{Multiple short lockdowns:} We wanted to evaluate the effect of shorter lockdowns at regular intervals instead of one long lockdown. Hence, the entire city (all sectors across all wards) were locked down on either i) The last two days of each week, or ii) The last one week of each month. The results are shown in Figure \ref{fig: multi}, and we find that weekend lockdown does work quite well. We also consider the possible effects of closing down only the education sector, and not locking down at all. It is found that the last scenario would have resulted in more than 10 times increase in the number of cases.
        \end{itemize}
        
    \item \textbf{Effect of reduced public transport usage:} This fraction was estimated to be about 0.17 in Kolkata prior to the pandemic. Hence, to account for lesser number of people opting for public transport, we see the effect of setting this fraction to [0.01, 0.05,  0.085, 0.1]. The results are shown in Figure \ref{fig: transport}   
    
    \item \textbf{Contact tracing efficacy:} To study the impact of changes in the efficacy of Contact Tracing, we varied the accuracy as follows: i) Workplace contact tracing efficacy - [60\%, 70\%, 80\%, 90\%, 100\%], and ii) Transport contact tracing efficacy - [30\%, 40\%, 50\%, 60\%, 70\%, 100\%] - indicating that those fractions of the contacts of each new positive case are traced. The results are shown in Figure \ref{fig: contact tracing}. 
    
    \item \textbf{Effect of compliance rate:} To evaluate the impact of having uniform compliance rates, we kept the Compliance Rate constant throughout the simulation, at [0.8, 0.7, 0.6, 0.5, 0.4, 0.3]. The results are shown in Figure \ref{fig: comp rate}
    
    \item \textbf{Effect of physical distancing:} We evaluated the effect of having different physical distancing factor (inter-person distance in the calculation of transmission probability). The inter-person distance was multiplied by a constant factor \textit{$SD_{factor}$} across all sectors. We evaluated the effect of setting \textit{$SD_{factor}$} as [2, 1.8, 1.6, 1.4, 1.2, 1, 0.8, 0.6, 0.4]. The results are shown in Figure \ref{fig: social dist}

\end{itemize}

In general, the results indicate clearly that the total number of cases can be controlled by high contact tracing, compliance rate and physical distancing. Prolonged lockdowns are unlikely to help, though lockdown on every weekend seems to be quite effective.

\section{CONCLUDING DISCUSSION}
In this paper, we discussed an agent-based model for simulation of Covid-19 pandemic in a city, that helps in assessment of intervention policies. The uniqueness of this model lies in the highly granular representation of the city including ward-level demographics, workplaces and healthcare facilities, as well as public transport usage and compliance rates etc. We also show that such granular representations are necessary for successful simulation. This model can work as a digital twin of a city, irrespective of the pandemic. Future improvements to this model will be to include vaccinations and economic impacts of the pandemic and policies. Another future goal is to use machine learning to estimate the parameters of the model. The most important usage of these models is their usage in "what-if" analysis. In this case, we explored the space of possible outcomes of different policies, and predicted possible results. Although it is impossible to confirm the accuracy of these results as they counterfactual, clearly they are consistent with common sense. A future direction of work will be to cast these in the framework of reinforcement learning and adaptively search the sequence of optimal policies.

\section*{ACKNOWLEDGMENTS}
This work was partially funded by Google Tensorflow Research Award, 2019 to IIT Kharagpur, and by the Department of Science and Technology, Government of India under its RAKSHAK scheme. We acknowledge the contributions of several students such as Shubham Mishra, Sparsh Kumar Jha, Sakshi Dwivedi and faculty members such as Profs. Sudeshna Sarkar, Debdoot Sheet and Arijit Mondal, all from IIT Kharagpur. We especially thank Prof. Bharath Aithal for providing us with the ward-wise datasets of Kolkata. 

\bibliographystyle{plain}
\bibliography{abmbib}

\end{document}